\documentstyle[aps,epsf,prl,preprint]{revtex}
\begin{document}
\draft
\title{Parametric resonance in atomic force microscopy: A new method
to study the tip-surface interaction}
\author{Franz-Josef Elmer}
\address{Institut f\"ur Physik, Universit\"at
   Basel, CH-4056 Basel, Switzerland}
\date{November 1997}
\maketitle
\begin{abstract}
We propose a new method to investigate interactions involved in
atomic force microscopy (AFM). It is a dynamical method relying on
the growth of oscillations via parametric resonance. With this method
the second and third derivatives of the tip-surface interaction
potential can be measured simultaneously. Because of its threshold
behavior parametric resonance AFM leads to sharp contrasts in surface
imaging.
\end{abstract}

\narrowtext
Atomic force microscopy (AFM) is a local method (on the scale from
nanometers to microns) able to sense the interaction potential
between a relatively flat surface and a more or less well-defined
probing tip mounted on a cantilever \cite{bin.86}.

The different methods to sense the interaction potential can be
divided into two groups: Static methods where the lever deflection
proportional to the force between tip and surface is measured
\cite{sar.91} and dynamic methods where the frequency shift of the
harmonically driven lever is measured \cite{sar.91}. From static
methods one can thus get the first derivative of the interaction
potential as a function of the tip-surface distance. With dynamic
methods where the oscillation amplitude is held small one gets the
force gradient, i.e., the second derivative of the interaction
potential by measuring the shift of the resonance frequency. There
are also dynamical methods with large oscillation amplitudes which
involve nonlinear oscillations induced by the interaction
\cite{zho.93,how.94,gie.95}. In all these dynamical methods the AFM
cantilever is excited near of one of its resonance frequencies.

In this Letter we propose a new dynamical method which is able to
measure the second and third derivative of the interaction potential
simultaneously. Contrary to the dynamical methods described above,
the AFM cantilever is excited near a frequency which is roughly {\em
twice\/} its resonance frequency. First we discuss qualitatively the
main properties of such a parametric resonance. In particular, we will
see from which kind of measurable data one can obtain the third
derivative of the potential. Following this qualitative discussion a
single-oscillator model is treated quantitatively. This leads to
convenient (but approximate) formulas for the second and third
derivative. Finally, we propose an operational scheme for measuring
the quantities from which one obtains the derivatives in question.

The reader might ask, can anything interesting happen by driving an
AFM cantilever away from its resonance frequency? The answer is: It
is possible to rapidly detect the growth of an instability due to the {\em
nonlinear\/} tip-surface interaction. The instability mechanism is
the first-order {\em parametric resonance\/} \cite{lan.60}.
Parametric resonance occurs in an oscillator which is driven by a {\em
modulation of its eigenfrequency\/} rather then by an external force.
The simplest example is a pendulum with an oscillating length. Such
oscillations can be excited very easily if the modulation frequency
is {\em twice\/} the average eigen frequency of the oscillator. Everybody
knows this intuitively from the experience of a swing-boat at a fair.

How does this mechanism manifest itself in an AFM cantilever which is
driven in the usual way? Upon approaching the sample, the resonance
frequency of the cantilever changes slightly by an amount
proportional to the second derivative of the interaction potential.
Thus the resonance frequency of the cantilever depends on the
tip-surface distance. If the cantilever is driven out of resonance, it
nevertheless oscillates with an amplitude which is of the same order
as the driving amplitude \cite{rem1}. Because of these oscillations
the tip-surface distance varies which then leads to a temporal
modulation of the resonance frequency. Thus parametric resonance can
take place. As an instability mechanism, parametric resonance is a
threshold phenomenon. The cantilever starts to oscillate only if the
modulation amplitude is larger than some threshold. Moreover, this
threshold depends on the ratio of the modulation frequency and the
eigen frequency. Its absolute minimum occurs when the modulation
frequency is twice the resonance frequency, the so-called first-order
parametric resonance condition. The minimal threshold is proportional
to the damping constant (i.e., the inverse quality factor)
\cite{rem2}.

For a fixed modulation amplitude above the minimal threshold and a
varying modulation frequency, one finds a frequency regime where parametric
resonance occurs. As shown below, the center of the interval is
determined by the second derivative of the interaction potential
whereas the third derivative determines the width of the interval. If
the modulation amplitude is much higher than the minimal threshold,
the effects of damping can be neglected. Then the maximum growth rate
(which occurs roughly in the middle of the instability interval) is
proportional to the third derivative of the potential. Because it
does not depend on the $Q$ factor, the instability may grow much
faster than one would expect from the slow relaxation in a system
with a high $Q$ factor.

\begin{figure}
\epsfxsize=120mm\epsffile{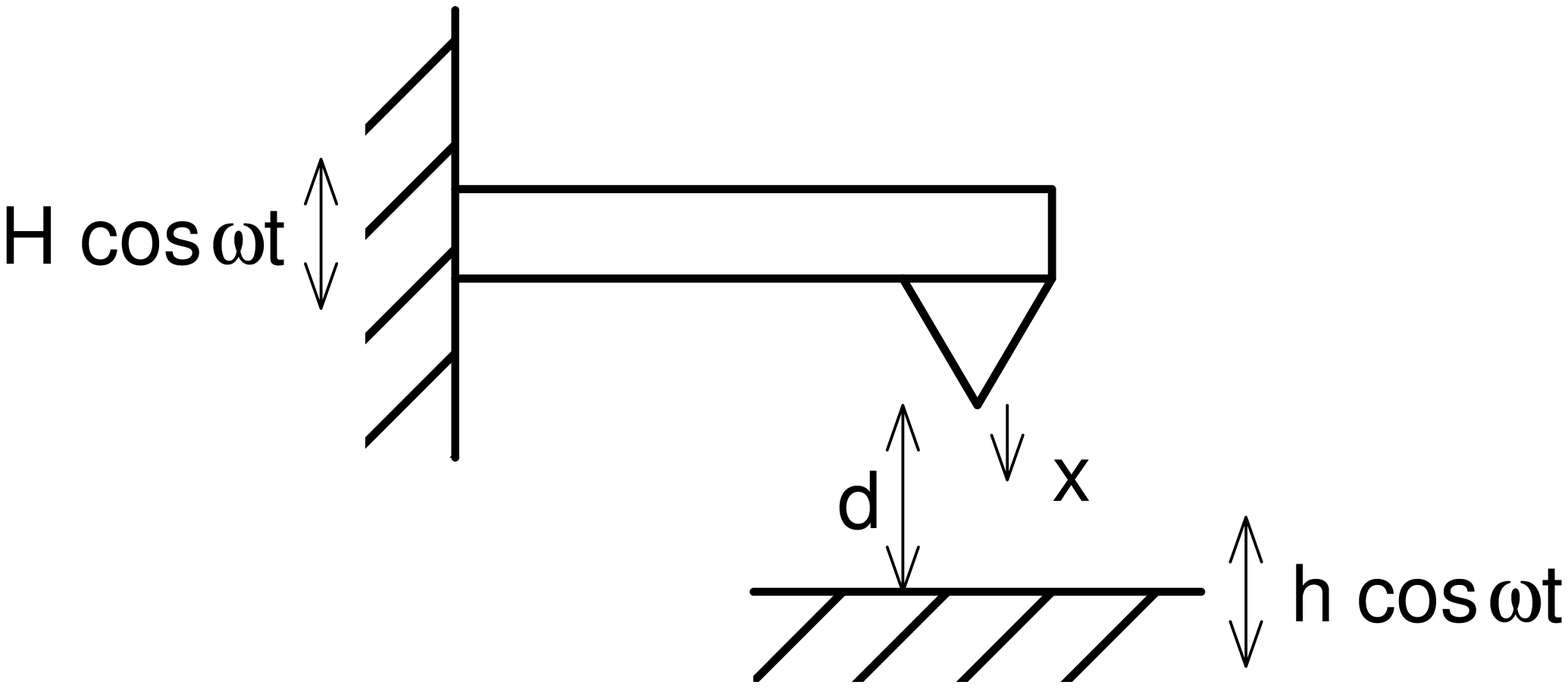}\vspace{3mm}
\caption[fsys]{\protect\label{f.sys}A schematic sketch of an atomic
force microscope (AFM). The tip-surface distance is either modulated
directly (modulation amplitude $h$) or indirectly by exciting an
oscillation in the cantilever (driving amplitude $H$).
}
\end{figure}

After this qualitative discussion we calculate parametric resonance
in an AFM by a model which describes the AFM by a single oscillator
(see also Fig.~\ref{f.sys}):
\begin{equation}
 \frac{\kappa}{\omega_0^2}\ddot x+\frac{\kappa}{\omega_0 Q}\dot x
   +\kappa x-V'(d+h\cos\omega t-x)=0,
  \label{eqm}
\end{equation}
where $\kappa$ is the spring constant, $\omega_0\equiv 2\pi f_0$ is
the resonance frequency, $x$ is the tip position relative to its
equilibrium value, $d$ is the tip-surface distance in equilibrium
without oscillations, $h$ and $\omega\equiv 2\pi f$ are the
modulation amplitude and frequency, respectively, 
and $V(r)$ is the tip-surface
interaction potential. Derivatives of $V$  are denoted by quotation
marks, e.g., $V'(r)\equiv dV(r)/dr$.

The tip-surface interaction potential has in general an attractive
and a repulsive part. In principle the proposed parametric resonance
method can work in non-contact mode as well as in contact mode. Here
we will restrict ourself to the non-contact mode (i.e., the tip is in
the attractive part of the potential). In the intermittent or
trapping mode analytical approach becomes difficult because tip and
sample deformations must be taken into account. Furthermore, the
damping is more complicated because it increases significantly upon
contact \cite{zho.93}.

In our model we use for the attractive part of the tip-surface
interaction the following effective potential \cite{isr.91}
describing the sum of Van der Waals interactions between a plane
sample and a tip with a spherical apex:
\begin{equation}
  V(r)=-\frac{AR}{6r},
  \label{V}
\end{equation}
where $A$ is the so-called Hamaker constant, $R$ is the tip radius,
and $r\ll R$ is the tip-surface distance. We have chosen $A=2.5\cdot
10^{-19}$J and $R=10$nm.

In (\ref{eqm}) we have assumed that the tip-surface distance is
modulated directly by vibrating the sample vertically. Alternatively
one can excite the cantilever at its support which also leads to an
oscillation of the tip-surface distance (see Fig.~\ref{f.sys}). Here
we restrict our discussion to the former case because in the single
oscillator model both kind of driving are mathematically equivalent.
That is, driving the support of the cantilever with the amplitude
$H=h\sqrt{[1-(\omega/\omega_0)^2]^2 +(Q^{-1}\omega/\omega_0)^2}$
leads to a tip-surface oscillation with amplitude $h$.

For $h\neq 0$ the equation of motion (\ref{eqm}) has a solution
$x_h(t)$ with the same periodicity as the modulation term, i.e.,
$x_h(t+2\pi/\omega)=x_h(t)$ for any $t$. Parametric resonance means
that the solution $x_h$ becomes unstable. That is, a tiny
perturbation $\delta x$ added to $x_h$ increases exponentially in
time. Linearizing (\ref{eqm}) in $\delta x$ and assuming
\begin{equation}
 x_h\ll h\ll d
  \label{con1}
\end{equation}
we get
\begin{equation}
 \frac{\kappa}{\omega_0^2}\ddot{\delta x}+\frac{\kappa}{\omega_0
Q}\dot{\delta x}+(x+V''_0+V'''_0h\cos \omega t)\delta x=0,
  \label{leqm}
\end{equation}
where
\begin{equation}
  V''_0\equiv \left.\frac{d^2V}{dr^2}\right|_d\quad{\rm and}\quad
  V'''_0\equiv \left.\frac{d^3V}{dr^3}\right|_d.
  \label{vss}
\end{equation}
This linearized equation of motion (\ref{leqm}) 
is the well-known Mathieu equation
with a damping term. The Floquet theorem implies that (\ref{leqm})
has two linearly independent solutions $c_1(\omega t)e^{\lambda_1t}$
and $c_2(\omega t)e^{\lambda_2t}$ with
$c_{1/2}(\phi+2\pi) =c_{1/2}(\phi)$. The so-called Floquet exponent
$\lambda$ is in general complex. Any solution of (\ref{leqm}) is a
superposition of these solutions. The solution $x_h$ is stable if any
disturbance $\delta x$ decays. The stability condition is therefore
Re$\lambda_{1/2}<0$.

\begin{figure}
\epsfxsize=120mm\epsffile{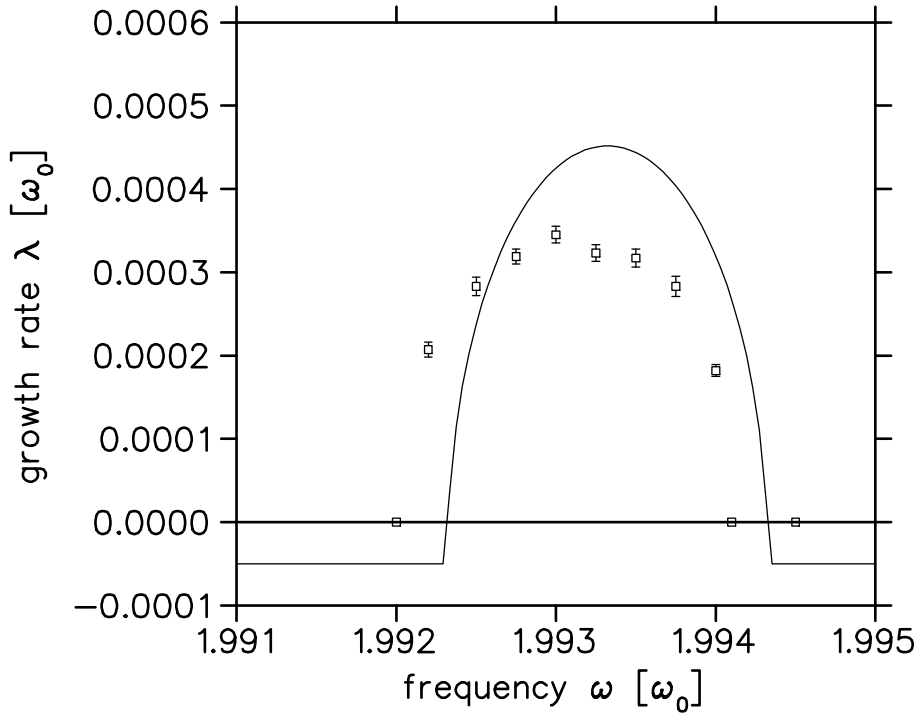}\vspace{3mm}
\caption[fgs]{\protect\label{f.gs}Growth rate $\lambda$ as a function
of the modulation frequency $\omega$. Frequencies and growth rates
are measured in units of the resonance frequency $\omega_0$ of the
cantilever far away from the surface. The solid line is the solution
of the characteristic polynomial (\ref{cp}). Squares show the result
obtained from simulations as shown in Fig.~\ref{f.onoff}. The
parameters are $d=5$nm, $\kappa=1$N/m, $Q=10^4$, and $h=0.5$nm.
}
\end{figure}

\begin{figure}
\epsfxsize=120mm\epsffile{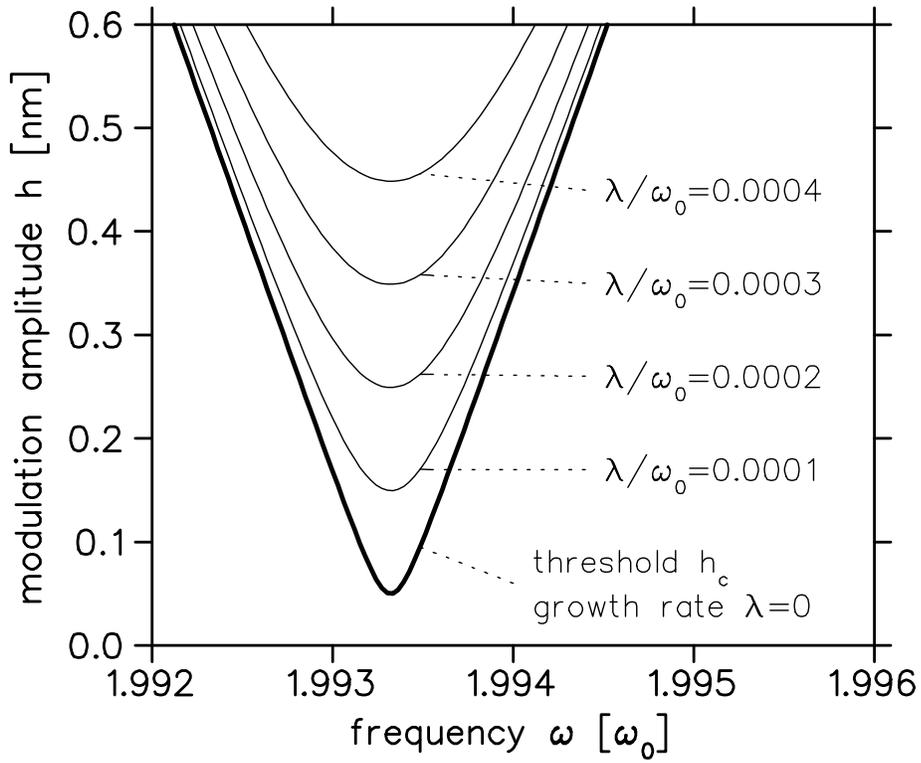}\vspace{3mm}
\caption[fit]{\protect\label{f.it}
Instability threshold $h_c$ and curves of constant growth rate
$\lambda$ as functions of the modulation frequency $\omega$. The
parameters are the same as in Fig.~\ref{f.gs}.
}
\end{figure}

In order to calculate the Floquet exponents one has to expand the $c$
functions into Fourier series. This expansion turns (\ref{leqm}) into
a set of infinitely many linear algebraic equations for the
coefficients of the Fourier series. In the case of weak damping and
weak driving, i.e.,
\begin{equation}
  Q\gg 1,\quad\mbox{and}\quad\frac{|V'''_0|h}{\kappa}\ll
   1+\frac{V''_0}{\kappa},
  \label{wd}
\end{equation}
one gets good approximations for the Floquet exponents by restricting
the Fourier series to the two leading components. That is, we solve 
(\ref{leqm}) approximatively with the ansatz
\begin{equation}
  \delta x=(a_+e^{i\omega t/2}+a_-e^{-i\omega t/2})e^{\lambda t}
   +\mbox{c.c}
  \label{ansatz}
\end{equation}
Plugging the ansatz into (\ref{leqm}) and neglecting terms with
$\exp(\pm 3i\omega t/2)$ leads to two linear homogeneous equations
for $a_\pm$ which have a nontrivial solution only if $\lambda$ is a
solution of the characteristic polynomial
\begin{eqnarray}
\lefteqn{\left(\frac{V'''_0h}{2\kappa}\right)^2-\left[\left(
   \frac{\lambda}{\omega_0}\right)^2+\frac{\lambda}{\omega_0Q}+1
   +\frac{V''_0}{\kappa}-\left(\frac{\omega}{2\omega_0}\right)^2
   \right]^2-}\nonumber\\
   &&\hspace{40mm}-\left(\frac{\lambda}{\omega_0^2}
   +\frac{\omega}{2\omega_0Q}\right)^2=0.
  \label{cp}
\end{eqnarray}
This equation can be solved analytically because it is a second order
polynomial in $(\lambda/\omega_0+1/2Q)^2$. Two solutions are always
complex with a negative real part. The two other solutions become
real if $h$ is large enough. One of them is always negative whereas
the other one becomes positives if
\begin{equation}
  h>h_c\equiv\frac{2\kappa}{|V'''_0|}\sqrt{\left[1+\frac{V''_0}{\kappa}
   -\left(\frac{\omega}{2\omega_0}\right)^2\right]^2+
   \left(\frac{\omega}{2\omega_0Q}\right)^2},
  \label{hc}
\end{equation}
where $h_c$ is the threshold for parametric resonance. Typical growth
rates given by the solution of (\ref{cp}) with the maximum real part
is shown in Fig.~\ref{f.gs}. The threshold curve $h_c$ and curves of
constant growth rates are shown in Fig~\ref{f.it}.

The minimum value of $h_c$ is given by $2Q^{-1}\kappa/|V'''_0|$ which
occurs at the parametric resonance condition
$\omega=2\omega_0\sqrt{1+V''_0/\kappa}$. Note that the minimum value
of $h_c$ increase like $d^4$. Thus for increasing distance $d$ the
minimum value of the modulation amplitude that is necessary to lead to
parametric resonance eventually exceeds the distance itself. Beyond
that point which scales like $Q^{1/3}$ parametric resonance is no
longer possible.

For $h\gg 2Q^{-1}\kappa/|V'''_0|$ the damping term in (\ref{cp}) and
(\ref{hc}) can be neglected. Using (\ref{hc}) one finds that
parametric resonance occurs for
\begin{equation}
  \omega_-<\omega<\omega_+,\quad\mbox{with}\quad\omega_\pm=
  2\omega_0\sqrt{1+\frac{V''_0}{\kappa}\pm\frac{|V'''_0|h}{2\kappa}}.
  \label{wpm}
\end{equation}
From measuring $\omega_\pm$ one obtains $V''_0$ and $V'''_0$. The
growth rate of a disturbance is zero at $\omega=\omega_\pm$ and has
its maximum near $2\omega_0\sqrt{1+V''_0/\kappa}$. The maximum growth
rate reads
\begin{equation}
  \lambda_{\rm max}=\frac{|V'''_0|h}{4\kappa}
   \sqrt{1+\frac{V''_0}{\kappa}}.
  \label{mgr}
\end{equation}
Note that the handy formulas (\ref{wpm}) and (\ref{mgr}) hold only if
the assumptions (\ref{con1}) and (\ref{wd}) are fulfilled. The
parameters of Figs.~\ref{f.gs} and \ref{f.it} clearly fulfill these
assumptions.

In order to measure the (positive) growth rate in parametric
resonance AFM we propose the following operation procedure. The
procedure switches between two modes of operation: The {\em on\/} and
the {\em off\/} mode. In the {\em on\/} mode the system is modulated
at roughly twice the resonance frequency. In the response of the
cantilever we are looking for oscillations at half of the driving
frequency (i.e., near the resonance frequency of the cantilever)
because such oscillations will be excited if parametric resonance
occur. If there is an instability the amplitude $a$ of these
oscillations increases exponentially and eventually exceeds some
predefined value $a_2$. Then the system switches into the {\em off\/}
mode where the modulation is switched off. The parametrically excited
oscillation decays until the oscillation amplitude decreases below a
value $a_1<a_2$ which is of the order of the noise level. Then the
system switches back into the {\em on\/} mode. By measuring the
average duration $t_{\rm on}$ of the {\em on\/} cycles one gets the
growth rate $\lambda$. Assuming that $x$ has to be increased from the
noise level $\sqrt{\langle x^2\rangle}$ up to $a_2$ we get
\begin{equation}
  \lambda\approx \frac{\ln\left(a_2/\sqrt{\langle x^2\rangle}\right)}
   {t_{\rm on}}.
  \label{lex}
\end{equation}

Because the quality factor $Q$ in vacuum is very high (i.e., of the
order of $10^4$) the decay of the oscillation would take much more
time than the excitation. In order to shorten $t_{\rm off}$ one can
use an external control circuit which brings the cantilever much
faster back to equilibrium.

\begin{figure}
\epsfxsize=120mm\epsffile{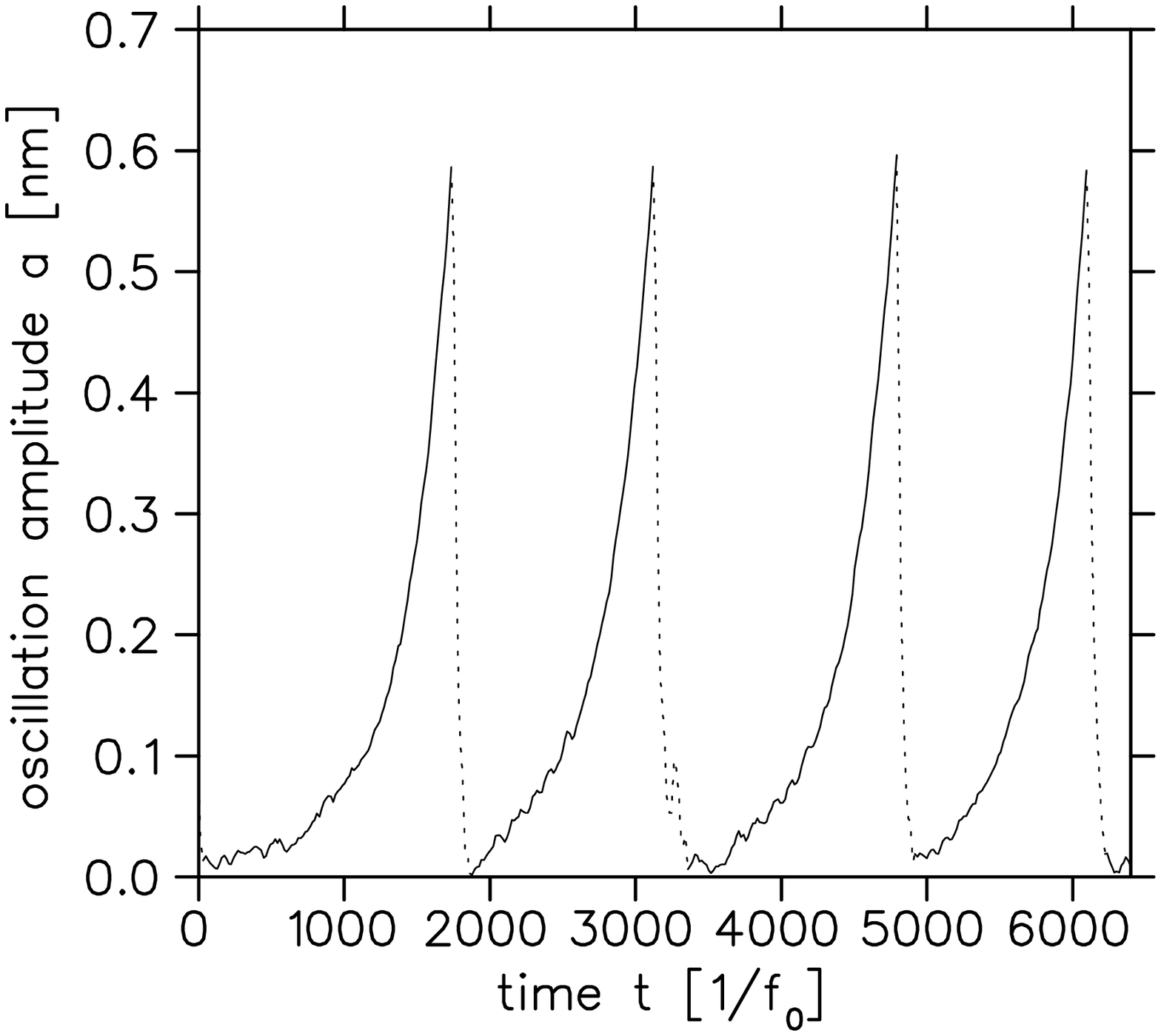}\vspace{3mm}
\caption[fonoff]{\protect\label{f.onoff}
Simulation of an AFM. Solid lines show the {\em on\/} mode where the
tip-surface distance is modulated with the frequency
$\omega=1.993\omega_0$. This modulation is switched off when the
amplitude $a$ [defined by (\ref{a})] reaches a threshold $a_2$ (here:
$a_2=0.6$). In the {\em off\/} mode (dotted lines) the quality
factor $Q$ is set to a relatively low value $Q_0$ (here: $Q_0=100$).
This simulates in a simplified way a control which brings the
oscillating AFM back to equilibrium on a fast time scale. When the
amplitude $a$ is below a critical value $a_1<a_2$ (here: $a_1=0.01$)
the system switches back to the {\em on\/} mode. The parameters are
the same as in Fig.~\ref{f.gs}. Noise level: $\langle
x^2\rangle=10^{-3}$.
}
\end{figure}

We have simulated this operational scheme by integrating the equation
of motion (\ref{eqm}) numerically. In order to incorporate the effect
of noise we have added on the right hand side of (\ref{eqm}) an
additive force
\begin{equation}
  \kappa\sqrt{\frac{2\langle x^2\rangle}{\omega_0Q}}\nu(t),
  \label{sf}
\end{equation}
with white noise, i.e.,
\begin{equation}
  \langle\nu(t)\rangle=0,\quad\langle\nu(t)\nu(t+\tau)\rangle
   =\delta(\tau).
  \label{noise}
\end{equation}
To extract the amplitude of the cantilever at half of the driving
frequency we used a simple band filter, i.e.,
\begin{equation}
  \ddot y+2\gamma\omega_0\dot y+\left(\frac{\omega}{2}\right)^2
   y=x.
  \label{filter}
\end{equation}
For the relative bandwidth we have chosen $\gamma=0.05$. From the
solutions of (\ref{filter}) we define an oscillation amplitude
\begin{equation}
 a\equiv\gamma\omega_0\sqrt{(\omega y)^2+(2\dot y)^2}.
  \label{a}
\end{equation}
For $x(t)=a_0\cos(\omega t/2)$ one gets exactly $a=a_0$. In our
simulation we have modeled the external control in the {\em off\/}
mode by adding on the right-hand side of (\ref{eqm}) the force
$-\kappa\dot x/(\omega_0Q_0)$. Figure~\ref{f.onoff} shows an example
of such a simulation.

In Fig.~\ref{f.gs} the growth rates obtained from such simulations
are compared with the analytical result. each square (with error
bars) are the average of 30 to 50 {\em on\/} cycles. There is a good
agreement between the analytical result and the simulation concerning
$\omega_\pm$. The growth rate from the simulation is in general to
small because Eq.~(\ref{lex}) assumes that the parametrically excited
mode has initially the amplitude $\sqrt{\langle x^2\rangle}$
neglecting the fact that in the noise there are also contributions
from the stable modes. Thus the initial amplitude of the
parametrically excited modes is on average less than
$\sqrt{\langle x^2\rangle}$.

With our simulations we have shown that parametric resonance AFM is
possible. Instability intervals and growth rates can be measured by
using our proposed on-off scheme. From this data one gets with the
help of (\ref{wpm}) and (\ref{mgr}) the second and the third
derivative of the tip-surface interaction potential. This method
should work very well in the non-contact mode. Using this method in
imaging we expect sharp contrasts because parametric resonance is a
threshold phenomena.

\acknowledgments
I am gratefully acknowledge helpful discussions with the group of Prof.
H.-J. G\"untherodt especially with M. Bammerlin, A. Baratoff, and E.
Meyer.

\newpage

\begin{references}
\bibitem{rem1}
We assume the absence of any higher eigen mode of the cantilever near
the driving frequency.
\bibitem{rem2}
In general there are also parametric resonances of higher order but
they have higher thresholds \cite{lan.60}.
\end{references}
\end{document}